\documentclass[11pt,a4paper]{article}
\usepackage{graphicx,hyperref,amsmath,natbib,bm,url}
\usepackage[utf8]{inputenc} %

\usepackage{color}
\definecolor{darkblue}{rgb}{0.15,0,0.37}
\definecolor{darkred}{rgb}{0.35,0,0.08}
\definecolor{mygrey}{rgb}{0.85,0.85,0.85} %

\hypersetup{
colorlinks=true,
hypertexnames=false,
colorlinks,
linkcolor={black},
citecolor={black},
urlcolor={black},
pdfstartview={FitV},
unicode,
breaklinks=true
}

\usepackage{amsthm}
\theoremstyle{definition}
\newtheorem{hyp}{Hypothesis}

\usepackage{microtype}
\usepackage[a4paper,text={14.5cm, 25.2cm},centering]{geometry} %
\usepackage{mathpazo} %

\usepackage{longtable, booktabs, tabularx} %
\usepackage{caption}  %
\usepackage[flushleft]{threeparttable}  %

\usepackage{multicol}
\usepackage{etoolbox}
\usepackage{relsize}
\patchcmd{\thebibliography}
  {\list}
  {\begin{multicols}{2}\smaller\list}
  {}
  {}
\appto{\endthebibliography}{\end{multicols}}

\usepackage[onehalfspacing]{setspace} %

\usepackage[marginal]{footmisc} %
\usepackage{pdflscape} %

\usepackage[capposition=top]{floatrow} %

\usepackage{color, colortbl}
\usepackage{multirow}

\newcommand\blfootnote[1]{%
  \begingroup
  \renewcommand\thefootnote{}\footnote{#1}%
  \addtocounter{footnote}{-1}%
  \endgroup
}

\usepackage{fancyhdr}
\usepackage{lastpage}
\usepackage{epigraph}

\begin{document}

\vspace*{0.5cm}

\thispagestyle{empty} %

\begin{center}
{\LARGE {The Impact of Acquisitions in the Biotechnology Sector on R\&D Productivity}}\\[0.4cm]
 {\large Luca Verginer${}^{1*}$, Massimo Riccaboni${}^{2}$}\\[0.2cm]
 {\small ${}^{1}$ Chair of Systems Design at ETH Zürich}\\
 {\small ${}^{2}$ IMT School for Advanced Studies Lucca}\\
\end{center}

\blfootnote{\textit{${}^*$ Corresponding author: lverginer@ethz.ch}}

\vspace*{-0.1cm}
\begin{center}
\begin{minipage}{0.9\textwidth}
\textbf{Abstract.}
This study examines the effects of acquisitions on the retention and R\&D productivity of inventors in the biotech sector, using data from 15,318 inventors involved in 1,375 acquisitions between 1990 and 2010. We employ a staggered difference-in-differences approach and find that acquisitions lead to a 13.5\% decrease in inventor retention and a 35\% drop in citation-weighted patent productivity post-acquisition. The productivity decline is more severe for inventors who remain with the acquiring firm, particularly for those whose expertise is closely tied to the target company. However, older inventors and those whose expertise aligns with the acquiring company's existing R\&D portfolio tend to retain higher productivity levels after the acquisition.

\\[0.3cm]
\begin{footnotesize}
\end{footnotesize}
\end{minipage}
\end{center}

\epigraph{The best people are staying, so I'm not super worried.}{\textit{Tweet by Elon Musk, 18th November 2022}}

\section{Introduction}\label{sec:introduction}
The impact of the acquisition of high-tech companies on innovation is still a matter of debate.
On one hand, acquisitions are beneficial as they provide start-up inventors with additional resources, thereby potentially increasing R\&D productivity.
On the other hand, acquisitions can reduce the productivity of inventors working in the target company.
This reduction in productivity can happen both unintentionally \citep{hitt1991effects,Kapoor2007} and intentionally, as in the case of 'killer acquisitions' \citep{cunningham2021}.

More generally, the innovation performance of a company after an acquisition depends on its ability to reorganize its activities in a way that leverages the assets and capabilities of the acquired company \citep{Hussinger2018} and on the ability of inventors to adapt to the new organizational environment or to find a viable exit strategy \citep{Kim2022}.

Crucially for R\&D-intensive industries, the acquisition's success depends on both tangible and intangible assets.
Therefore, intellectual property, both codified (e.g. patents) and tacit (i.e. crucial employee know-how), must be managed carefully so as not to jeopardize acquisitions.
High uncertainty and asymmetric information between the acquiring company and the target company's inventors can lead to disagreements about the potential value of their knowledge in the new organizational environment. The turnover of inventors after acquisitions is a consequence of such disagreements.

This is particularly relevant in knowledge-intensive industries after a paradigmatic change, such as in the life sciences, which saw a rush of M\&A at the turn of the century \citep{riccaboni2002firm} to absorb knowledge in the emerging field of biotechnology.\footnote{M\&A deals in 1999 involving US companies were worth well over 500 Billion USD \citep{Danzon2007}, highlighting the economic importance of the practice.}

At that time, the primary of these acquisitions were new biotechnology firms (NBFs) that had been founded since the mid-1970s and were poised to challenge the leadership of the established pharmaceutical companies \citep{orsenigo2001technological, riccaboni2002firm, beraud2020, pammolli2021}.
The biopharmaceutical sector is an exceptionally R\&D intensive sector, with \citet{PRMA2012} claiming that up to 17\% of revenue is spent on R\&D, compared to the US industry average of 4\% \citep{Danzon2007}.
The importance of patents in particular and intellectual property in general is well known in the biopharmaceutical industry \citep{cohen2000}. Therefore, the knowledge embodied in the teams of inventors is crucial for developing innovative drugs.

Inventors, the employees behind these innovations, play a crucial role in knowledge production; therefore, their innovation productivity after M\&As is vital.
However, there are various reasons why an inventor may be less productive after an acquisition, including resignation or dismissal, which can temporarily impede R\&D productivity.
These setbacks may be due to restructuring, cost-cutting measures, a change in the company's strategic direction or even killer acquisitions.
In these cases, the departure of an inventor may not be voluntary but rather a consequence of the post-acquisition environment and company decisions.

Acquired inventors \citep{kim2020startup} may also experience a radical company restructuring with a reconfiguration of R\&D teams that could affect innovation.
The departure of some inventors are critical, especially considering their tacit and 'socially complex' knowledge \citep{Barney1991} --- firm capabilities emerging from complex social interactions. 
Moreover, the existing literature on the success of M\&As shows that technological complementarities are important for post-merger innovation productivity.
Nevertheless, it is crucial to understand the response of inventors to voluntary or involuntary M\&As.

Inventors who work for an acquired company can react to the acquisition in different ways: they can (1) stay with the company and continue to contribute to innovation and patent production. However, this is not the only alternative they have. The alternatives we focus on in this paper are that they (2) continue their patent activity, but for a third party unrelated to the acquisition, or (3) switch to a different role in which former inventors are no longer involved in patent production.

Most of the extant literature addresses issues of turnover and performance in the context of top management (e.g. \citet{Hambrick1993,Haveman1995,Haleblian2009}), general employee turnover \citep{Carriquiry2017,arnold2023job} and, to a lesser extent, the impact of M\&A on R\&D personnel \citep{Ernst2000,Hussinger2018}. %
Most of the literature documented a productivity loss after M\&As \citep{hitt1991effects,paruchuri2006,Kapoor2007,ornaghi2009mergers,cunningham2021}.
However, one of the main shortcomings of the existing literature on the relationship between the acquisitions of innovative firms and the productivity of inventors is the possible presence of confounding factors that could increase the likelihood of a firm becoming an acquisition target, inventor turnover, and the likelihood of inventors becoming less productive. Therefore, it is difficult to establish a causal relationship between the event (i.e. the takeover) and the productivity of investors.

In our analysis, we rely on a staggered difference-in-differences methodology to overcome this limitation.
To better isolate the consequences of integration on R\&D output, we focus on a specific period of the life cycle of an innovation-intensive industry immediately after the molecular biology revolution in the pharmaceutical industry.

We look specifically at acquisitions in the biotechnology industry for two reasons.
First, it is an innovation-intensive sector and therefore relevant for our analysis of the retention of skilled human capital and intellectual property.
Second, the pharmaceutical industry experienced an increased rate of M\&A events between 1990 and 2010, ranging from mergers between large pharmaceutical companies to acquisitions of new biotechnology companies.
During the same period, \citep{riccaboni2002firm} and \citet{tomasello2017rise} found that research alliances in general and the biotechnology industry in particular were on the rise. This finding shows that companies were eager to pool their resources to be at the forefront of R\&D and that M\&A is a natural next step in pooling resources.
Focusing on the first wave of acquisitions of biotech companies also gives us the opportunity to examine the long-term impact of these events on patent production.

In our analysis, we address the question of whether an acquisition affects inventors' productivity.
Note that we \emph{only} consider outright acquisitions and not any weaker forms of M\&A such as partial or full mergers. The treated units of analysis covered are firms and inventors involved in acquisitions. We use only acquisitions to avoid some of the difficulties we might encounter when considering mergers.\footnote{See \citet{ornaghi2009mergers} for a similar analysis on the impact of mergers on innovation in the pharmaceutical sector.}
First, by looking only at acquisitions, it is clear which firm will have control and that no new legal entity emerges from the transaction.
Secondly, the relationship between the acquired and acquiring company allows us to identify the target of the transaction that is most likely to be affected by changes, i.e., the acquired firm.

In our analysis we aim at estimating the turnover rate due to takeovers, which we find to be 13.5\% in the five years after the event, in line with \citet{Ernst2000} and \citet{Ranft2000}. We also analyze the causal impact of takeovers on inventors' patent production and find that the citation-weighted number of patents decreased by 35\%. We show that more inventors stop producing patents after takeovers. In general, inventors who remain with the acquiring firm suffer a larger loss, but the effect is mitigated by the inventors' experience, technological overlap and lower technological specificity to the target firm's knowledge. 

Our paper complements the literature on the consequences of a firm's acquisition on R\&D productivity by providing compelling evidence that M\&As have a detrimental effect on R\&D productivity and how inventors can better cope with these traumatic events.

\section{Background and Hypotheses}\label{sec:lit}
In this paper, we take a knowledge-based view of the firm. According to \citet{Grant1996}, a firm is a vehicle that organises tacit and complex social knowledge of individuals to create products and services. to argue that among the various motives for acquiring innovative firms, in addition to the acquisition of codified intellectual property (i.e. patents), the retention of employees' tacit knowledge is a key factor in R\&D-intensive industries. However, it is well known that acquisitions have a disruptive effect on the acquired workforce \citep{Kapoor2007,arnold2023job}.

Takeovers could mean a dramatic change in the organization of R\&D of the target company and a reduction in the autonomy and independence of inventors who may prefer to leave the company.
\citet{Hambrick1993} finds evidence that the departure of key employees was a significant predictor of poor post-merger performance.
The authoritative work of \citet{Coff1997} on the management of human resources and their tacit knowledge reinforces this view. He argues that the retention of key employees is a major challenge in takeovers. \citet{Carriquiry2017} find that the disruptive nature of an acquisition can have an overall negative effect on turnover rates.

Several studies focusing on M\&As in the Nineties have found that top executives and management are more likely to leave the firm following an acquisition. %
\citet{Ranft2000} finds that according to a survey involving 89 firms which were part of M\&As senior executives were the most likely to leave and 22.7\% R\&D personnel left on average. However, they also find that R\& D personnel is considered to be the most crucial class of employees to retain. This effect is exacerbated when the target of the acquisition are high-tech startup companies \citep{Kim2022} like in the case of biotech firms. A result of this literature is that acquisitions affect retention rates.

Specifically, due to changes in routines and managerial hierarchies as well as uncertain career prospects \citep{Hobman2004}, an inventor may decide to leave a company. Similarly, \citet{Holtom2005} argue that shocks such as M\&A `trigger' a reevaluation of career and life goals of the affected employee. For example, an employee wanting to leave all along takes the M\&A as a sign to act, or similarly, differences in the managerial ``culture'' could trigger a reevaluation.

It emerges clearly from the meta-analysis on the turnover by \citet{Griffeth2000} that the motivations for turnover are many and varied. Still, shocks and disruption to business-as-usual can lead to higher than average turnover.

The departure of inventors following an acquisition often leads to reduced R\&D productivity. Inventors typically collaborate in teams, and reconfiguring these teams post-acquisition is a time-consuming process that can negatively affect individual productivity. Departing inventors face the challenge of integrating into new teams, while those who remain may have to adapt to disrupted routines and struggle to collaborate effectively within the new organizational structure, often feeling the absence of their former colleagues' contributions.

\citet{Kapoor2007} observed a lasting negative impact on inventor productivity in the semiconductor industry due to acquisitions. However, \citet{Kapoor2007} also notes, few studies have focused on the individual inventors’ perspective, largely because tracking their long-term performance is challenging.

These observations and results lead us to formulate Hypothesis~\ref{hyp:leave}:%

\begin{hyp}[H\ref{hyp:leave}]\label{hyp:leave}
    Acquisition are traumatic events with a negative effect on inventors' R\&D productivity. 
\end{hyp}

Some of these effects are unintended consequences, as the acquiring company may be interested in retaining only the most productive inventors. However, in the case of a killer acquisition, the acquiring companies pursue opposite goals \citep{cunningham2021}.

In addition, some of the best employees may voluntarily decide to leave the acquiring company \citep{arnold2023job}. More generally, \citet{kaul2021} identify some of the most important factors for potential disagreements between inventors and acquiring companies about the future potential of their ideas. According to the theoretical framework of \citet{kaul2021}, when there is great uncertainty about the potential future productivity of the target company's inventors, it is more likely that the inventors will leave the acquiring company, as they may have more problems securing access to resources in the new organization \cite{Kim2022}.

This is typically the case for young and less experienced inventors who do not yet have a record of producing patents, making it difficult for the acquiring firm to easily assess their innovation potential. Therefore, we can postulate the following research hypothesis.

\begin{hyp}[H\ref{hyp:young}]\label{hyp:young}
   Less experienced inventors are more likely to leave the acquiring company, and those who stay will suffer a greater loss of productivity.
\end{hyp}

A second important source of potential disagreement between acquired inventors and the acquiring company concerns the specificity of the inventor's target company specificity. If an inventor's ideas are based on knowledge that is unique to the target company's knowledge base, it is more difficult for the acquiring company to properly assess the potential value of his or her ideas. Less exclusive acquired inventors who work with other companies have more opportunities outside the company and are generally more likely to leave the company.

After an acquisition, the evaluation of potential ideas from specific inventors at the target company becomes more challenging for the acquiring firm. Additionally, the inventors who remain may face difficulties adapting to the new organizational structure. This effect is particularly pronounced in the case of acquisitions involving new biotechnology firms, as these start-ups emerged following a technological revolution in molecular biology. Based on these considerations, we put forward the following hypothesis,

\begin{hyp}[H\ref{hyp:young}]\label{hyp:exclusive}
   Exclusive inventors are more likely to leave the acquiring company, and those who stay will suffer a greater R\&D productivity loss.
\end{hyp}

Common reasons proposed for why M\&A take place, combine elements of vertical and horizontal integration, economies of scale and scope, and transfer of specific assets or capabilities \citep{Danzon2007, Higgins2006, ravenscraft2000paths}.
\citet{Higgins2006}, analysing pharmaceutical firms in the Nineties, find that companies with expiring patents and declining internal productivity are more likely to engage in M\&A.

The rationale behind this proposed mechanism is that expiring patents free up production and R\&D capacity.
In contrast, fixed and sunk cost assets (i.e.\ employees and factories) can be used to enhance the impact of an acquired company that has a viable compound but lacks productive and marketing capabilities.

Similarly, \citet{Villalonga2005}, looking at 9\,276 acquisitions between 1990 and 2000, find that the acquisition route, as opposed to alliances and divestitures, is chosen more often if the target has more valuable technological resources at its disposal.

Moreover, M\&A may lead to positive stock performance as \citet{Higgins2006} find.
M\&A identified as being conducted with the expressed purpose of R\&D consolidation, lead to abnormal returns to both the acquired and the acquiring company.
However, this comes with the caveat that the acquired firms were found more likely to experience financial difficulties \citep{Danzon2007}. In particular, \citet{ravenscraft2000paths} find that acquired firms tend to experience negative stock returns up to 18 months before the takeover.

We argued that the retention of inventors is essential and looked at possible reasons a given inventor might leave; however, the employment relationship will only continue if it is also in the best interest of the acquiring firm.

Possible reasons for terminating inventors after acquisitions vary, starting with duplicating skills already covered by the acquiring firm or abandoning a research area altogether. This last point is in line with the finding by \citet{Zhu2018} that acquiring firms tend to sell off acquired patents right after the event. \citet{ornaghi2009mergers} presents similar negative consequences for R\&D productivity for mergers in the pharmaceutical sectors.

However, it is still a valid assumption that the acquiring firm would not want to decimate its newly acquired inventors immediately, arguably valuable assets, but transition them out gradually. On the other hand, as argued above, inventors might see the shock as a nudge to change employer. Arguably, this can be more relevant for inventors with no technological overlap with the acquiring firm since their knowledge might be more valuable on the market than in the merged company. %

Concerning the duplication of expertise, technological similarity, and post-merger performance of firms, \citet{Kapoor2007}, \citet{ahuja2001technological}, \citet{cloodt2006mergers} find that the patenting output and innovativeness is highest if the technological profiles of the acquiring and acquired firm overlap.
In the case of biotech acquisitions the level of overlap is likely not to be critical, like in the case of big pharma mergers \citep{ornaghi2009mergers}. 
Moreover, the acquiring companies can better evaluate the potential of new ideas when the inventors are active in the same patent classes.

Following this line of reasoning, we should expect that inventors with a technological overlap in their technological profile with the acquiring firm are more likely to stay and and remain productive after an acquisition.

\begin{hyp}[H\ref{hyp:overlap1}]\label{hyp:overlap1}
    Inventors with higher technological overlap with the acquiring firm, are more likely to stay on and remain productive.
\end{hyp}

\section{Data}\label{sec:data}
For our analysis, we require two types of information: (1) detailed of acquisition events and (2) a comprehensive patent dataset containing details of assignees, inventors and citations.

\begin{figure}[!ht]
    \centering
    \includegraphics[width=1\textwidth]{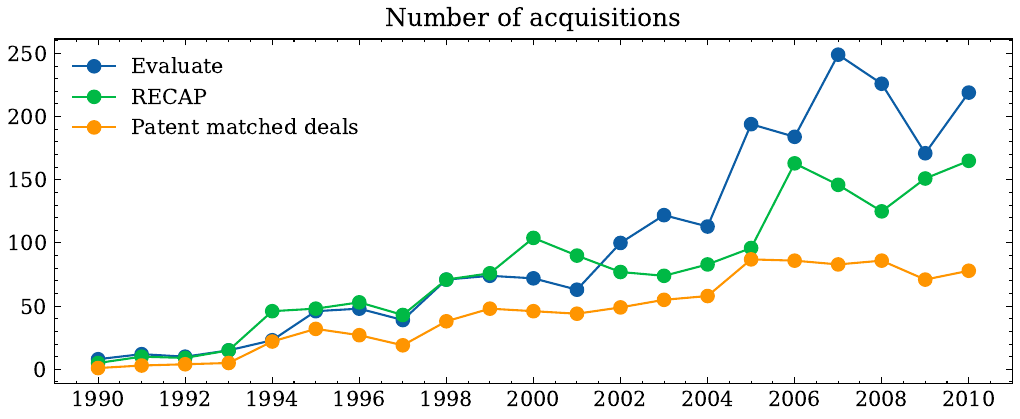}
    \caption[Annual deals in RECAP]{Number of acquisitions in the biotech industry as recorded in RECAP and EVALUATE. These datasets were merged based on the names of the companies involved and the year of the acquisition. Inclusion criteria for the analysis were that the acquired company had filed at least one patent before the acquisition and that the acquirer had filed at least one patent before and after the acquisition. The orange curve labeled ``Patent Matched Deals'' shows the subset of deals that met these criteria and were included in our study.}\label{fig:annual_repac_deals}
\end{figure}

Our study on inventor turnover in biotechnology is based on acquisition data from Thomson Reuters Recap\footnote{Now Cortellis Deals Intelligence, http://recap.com/} and Evaluate\footnote{https://www.evaluate.com/}.
These databases provide a comprehensive record of significant acquisition events in the industry.
Figure~\ref{fig:annual_repac_deals} shows the evolution of these acquisitions over time and highlighting an increase from a few acquisitions in 1990 to a peak of around 250 in 2007.
There are overlaps in the transactions listed in RECAP and Evaluate.
To ensure accuracy and eliminate redundancies, we merged acquisitions across the two datasets by matching on the names of the two firms and the year of the transaction.

Our analysis focuses on a specific subset of these transactions.
A prerequisite for inclusion in our dataset was that the acquired company had filed at least one patent application in the five years prior to the acquisition and that the acquired company had done so both before and after the event.
This was necessary to identify inventors who worked for the acquired company and to ensure that the company continued its R\&D activities after the acquisition.
After filtering based on these criteria, we identified 1,375 acquisitions.

To construct our counterfactual, we include in our study only firms that have been acquired, rather than firms that have not experienced such events.
Specifically, we apply a staggered difference-in-differences approach in which we compare companies that have already been acquired (treated) with those that are likely to be acquired in the future (not yet treated). This methodology, inspired by \citet{Kim2022}, has recently been used to analyze the impact of acquisitions on employee entrepreneurship.
The rationale for this approach is based on the premise that firms involved in acquisitions may have particular characteristics compared to those not acquired, possibly due to financial distress or the value of their intellectual property. Therefore, our analysis also includes companies that were not acquired during our observation period but could potentially be acquired in the future.
We assess the impact of acquisitions by comparing these `not yet acquired' inventors of companies with the acquired inventors.

For patent details, we rely on the disambiguated patent dataset from \citet{Morrison2017}, which provides a comprehensive set of patents filed with the USPTO, EPO and PCT, covering patents in the US, Europe and Japan. We rely on Google Patents for updated citation data through 2022.

\begin{figure}[ht]
    \centering
    \includegraphics[width=0.55\textwidth]{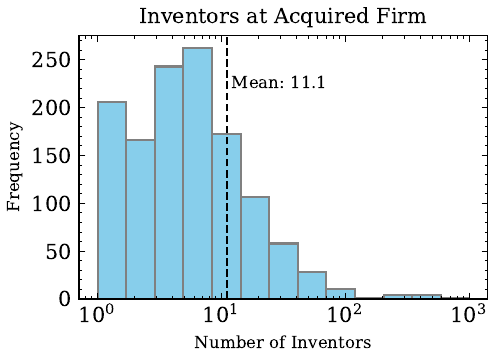}
    \caption{Histogram of unique inventors filing patents for the acquired company in the 5 years before the event. These correspond to the inventors in used in the main analysis.}
    \label{fig:inventors-per-deal}
\end{figure}

To identify the inventors of the acquired companies, we collect patents filed by these companies within five years prior to the acquisition and list the inventors therein. The inventors' names have been disambiguated so that we can track the inventors' patent activities both with the company and individually, which gives us a comprehensive picture of their patent activities and the firms they worked for.
We identified 15\,318, with an average of 11.1 inventors working for an acquired company (see Figure~\ref{fig:inventors-per-deal}).

Our analysis employs a staggered difference-in-differences approach to assess the impact of acquisitions on inventor outcomes, focusing on inventor retention, patenting activity, and citation-weighted patenting.
At the core of this analysis is the \textit{Deal Year} marking the timing of each acquisition.

We focus on four key outcomes: \textit{Left}, \textit{R\&D Activity}, \textit{Patents}, and \textit{Citations}. The first two, binary variables measure the impact of the acquisition on employee retention and inventors' ongoing R\&D activity.
\textit{Left} indicates whether the inventor has filed any patent applications after this year either for the acquiring no the acquired form.
\textit{No Longer Active} indicates inventors who filed their last patent in the previous year.

\textit{Patents} and \textit{Citations} measure productivity annually.
\textit{Patents} counts the number of patents the inventor has filed in a given year, and \textit{Citations} reflects the number of citations these patents have received until 2022.
We consider 12 years to be sufficient to assess the impact of the latest patents in our dataset (2010).

We consider several individual characteristics, including \textit{Age}, \textit{Tenure}, \textit{Exclusivity}, and \textit{Common IPC}. All variables are listed in Table \ref{tab:variables}. 
\textit{Age} is the number of years since the inventor's first patent filing, indicating their experience up to the point of acquisition.
\textit{Tenure} measures the number of years since the first patent the inventor filed for the acquired firm.
\textit{Exclusivity} is a central metric assessing the inventor's focus on the acquired firm. It is calculated as the ratio of patents filed for the acquired firm to their total patents in the five years preceding the acquisition.
Lastly, \textit{Common IPC} quantifies the technological alignment between the inventor and the acquiring firm.
It is computed as the number of International Patent Classification (IPC) classes shared between the inventor's patents and those of the acquiring firm on patents before the acquisition.
This overlap indicates the technological synergy between the inventor's expertise and the acquiring firm’s existing technological profile.

\begin{table}[h]
    \small
    \centering
        \caption{Detailed summary of the variables used in the analysis}
    \label{tab:variables}
        \begin{tabular}{lp{11.2cm}}
        \toprule
            \textbf{Variable} & \textbf{Description} \\
            \midrule
                 Deal Year$_{i}$ & Year in which the takeover was announced. \\
                 After$_{it}$ & Dummy variable: 1 if the $t$ is after the deal year, 0 otherwise.\\
                 Left$_{it}$ & Dummy variable: 1 if the last year $i$ filed a patent for either the acquiring or acquired company was before $t$, 0 otherwise. \\ 
                 Patents$_{it}$ & Number of patents filed by inventor $i$ in year $t$. \\
                 Citations$_{it}$ & Citations received by 2022 for patents filed by inventor $i$ in year $t$. \\
                 Age$_{i}$ & Professional age at the time of acquisition, measured as the number of years since the first patent application. \\
                 Tenure$_{i}$ & Years from the inventor's first patent filing for the acquired company to the year of acquisition. \\
                 Exclusivity$_{i}$ & Proportion of patents filed by the inventor for the acquired company in relation to her/his total patents in the five years prior to the acquisition. \\
                 Common IPC$_{i}$ & Number of shared International Patent Classification (IPC) subclasses between the inventor's patents and those of the acquiring firm before the acquisition. In our analysis, we use this variable in two forms: as a continuous measure and as a discretized variable categorized into three levels: 0, 1--3, and 4 or more.\\
                 R\&D Activity$_{it}$ & File at least on patent on or after year $t$. \\
        \bottomrule
    \end{tabular}
\end{table}

Tables \ref{tab:descriptives} and \ref{tab:corr} in the appendix contain some descriptive statistics and the correlation table for the variables we use in our analysis.

\section{Methodology}\label{sec:methods}

\subsection*{Difference-in-Difference with with multiple time periods}

To assess the treatment effect of acquisitions on inventor retention, continued patenting, and patent output, we use the group average treatment effect framework introduced by \citet{Callaway2021,rios2022drdid}.
The central concept of the approach is to analyze the causal effects of acquisitions by comparing units that have been treated with those that are awaiting treatment.
This method takes advantage of the different timing of acquisitions and allows us to assess the treatment effect on the groups before and after their treatment.
Specifically, we estimate the treatment effect for each time interval (e.g. one year) by comparing the outcomes of the units already treated with those yet to be treated.

\begin{align}\label{eq:event}
    Y_{ict} & =
    \sum^{T=5}_{\tau=-5} \lambda_\tau \text{After}_{ict\tau} + \alpha_i + \xi_{ct} + \epsilon_i
\end{align}

In this model, the dependent variable $Y_{ict}$ represents outcomes for inventor $i$ in company $c$ at time $t$.
The analysis spans a period beginning five years before the acquisition ($\tau=-5$) and ending five years after the acquisition ($T=5$).
The term $\sum^{T=5}_{\tau=-5} \lambda_\tau$ captures the time-varying effect of the acquisition on these inventor outcomes.

The key independent variable, $\text{After}_{c \tau}$, is a binary variable that indicates whether an acquisition has occurred for company $c$ at time $\tau$.
To account for unobserved time-invariant characteristics of inventors, the model incorporates individual fixed effects ($\alpha_i$).
Company and Year fixed effects ($\xi_{ct}$) are included to control for unobserved factors specific to company $c$ at time $t$ that may influence the inventor outcomes.
The error term ($\epsilon_i$) represents any unobserved factors not accounted for in the model that might affect the dependent variable.

Our main interest is in the coefficients $\lambda_\tau$ that estimate the effect of the takeover on inventors' outcomes for each year within the given period (5 years before to 5 years after the takeover).
This model aims to improve our understanding of the impact of acquisitions on inventor retention, the number of patents filed, and the citation-weighted number of patents filed while controlling for individual and firm-time fixed effects.

The analysis was performed using the \texttt{did} R package~\citet{did}, which implements the generalised diff-in-diff framework of \citet{Callaway2021}
The aggregate effects are shown in Figure~\ref{fig:ATT}, in which we summarise and average the Average Treatment Effect on the Treated (ATT) by treatment year groups and average them.
The overall ATT, as shown in Table~\ref{tab:att}, reflects the average effect across all periods.

\subsection*{Aggregate Interaction Design}

To streamline our analysis and estimate the average effect over five years after the event, we change the equation \eqref{eq:event} by estimating a single `After' parameter, $\lambda$, instead of the previous 11 different $\lambda_\tau$ parameters.
This simplification helps tackle the model's complexity, especially when additional interaction effects are included.
The revised model is:

\begin{align}\label{eq:pre-post1}
    Y_{ict} & = \lambda \text{After}_{ict} + \alpha_i + \xi_{ct} + \epsilon_i
\end{align}

Extending this model, we introduce interaction terms to assess the impact of factors such as inventor exit, age, exclusivity and technological overlap.
To examine the impact on patenting activity post-departure, we use:

\begin{align}\label{eq:pre-post2}
    Y_{ict} & = \lambda \text{After}_{ict} + \beta_1 \text{Left}_{i} + \beta_2 \text{After}_{ict} \times \text{Left}_{i} + \xi_{ct} + \epsilon_i
\end{align}

We omit individual fixed effects to avoid co-linearity with fixed inventor characteristics while retaining firm and time-fixed effects.
Here, $\beta_2$ represents the net effect of `Left' after acquisition on the outcome $Y_{ict}$.

By integrating interactions with other variables, we can further disaggregate the nuances of the takeover, focusing on the \textit{left} status.

\begin{align}\label{eq:exclusivity}
    Y_{ict} & =
    \lambda \text{After}_{ict} + 
    \beta_1 \text{Left}_{i} \\
    &+\beta_2 \text{After}_{ict} \times \text{Left}_{i} \nonumber \\
    &+\beta_3 \text{Exclusivity}_{i} \nonumber \\
    &+\beta_4 \text{Exclusivity}_{i} \times \text{After}_{ict}\nonumber \\
    &+\beta_5 \text{Exclusivity}_{i} \times \text{Left}_{i}\nonumber \\
    &+\beta_6 \text{Exclusivity}_{i} \times \text{Left}_{i} \times \text{After}_{ict}\nonumber \\
    &+\beta_7 \log(\text{Age}_{i}) + \beta_8 \log(\text{Tenure}_{i}) + \beta_8 \text{Common IPCs}_{i}\nonumber\\
    &+\xi_{ct} + \epsilon_i \nonumber
\end{align}

To understand the impact of exclusivity on inventors' post-acquisition productivity, we mainly focus on $\beta_6$, which captures the net effect of post-acquisition exit through the level of exclusivity.

To investigate the influence of age, we replaced \textit{Exclusivity} in \eqref{eq:exclusivity} with \textit{Log(Age)}.
We categorized $\text{Common IPCs}_{i}$ into three levels: 0, 1-3 and 4 or more to assess technological similarity.
By discretizing $\text{Common IPCs}_{i}$ and including these levels in our model, we can identify non-linear effects.

As an alternative approach, a matched diff-in-diff regression framework has been used in the literature, comparing treated inventors before and after acquisitions with comparable, never-treated inventors who worked in similar companies that never experienced an acquisition \citep{verginer2022impact}. However, this approach has two significant limitations. First, the company-level matching procedure means a significant reduction in the sample of acquisitions considered from more than a thousand to less than a hundred. Second, the untreated companies might differ from the target companies, and these differences might be unobservable. For these reasons, we prefer to compare treated inventors with not-yet-treated inventors working for firms that will be acquired later.

Another potential methodological problem concerns the observability of inventors. The activities of inventors can be tracked as long as they continue to file patents, either for the acquiring company or with different assignees. If the chance that inventors continue to file patents depends on the employer, the productivity difference between those who leave and those who stay might depend on the chance that they do not file their patents after the acquisition. However, using a Heckman regression to control for the probability that an inventor stops filing patents, it was found that the estimated probability that an inventor leaves the acquired company does not change, suggesting that the chance of \textit{going dark} does not affect the productivity of active inventors in different organizational settings \citep{verginer2022impact}. Therefore, in the following analysis, we consider the moderating effect of inventor characteristics on the productivity of active investors before and after takeovers.

\section{Results}\label{sec:results}

\subsection*{The impact of acquisitions on R\&D productivity}

Takeovers of high-tech companies are notoriously disruptive events, especially in terms of the turnover of inventors and other highly skilled workers, as noted by \citep{Hussinger2018}.
However, the impact of these takeovers goes far beyond the mere attrition rate.

\citep{Kapoor2007} found that acquisitions negatively affect the patent production of inventors working for semiconductor companies.
More recently, \citep{Kim2022} has shown that acquisitions significantly increase the rate of inventor entrepreneurship.
These studies illustrate the different consequences of takeovers in high-tech sectors.

In our analysis, we take a broader perspective to examine the causal effects of acquisitions on the mobility and productivity of inventors who worked at the target firm before the acquisition.
We consider the impact of acquisitions in four dimensions:
(i) the turnover of inventors from acquired firms;
ii) the innovation productivity of inventors from acquired firms, measured by the number of patents filed;
iii) the citation-weighted number of patents of inventors from the target company as an alternative measure of innovation productivity; and
iv) the probability that inventors will no longer apply for patents after the acquisition.

To quantify these effects, we use a staggered difference-in-differences approach, which is explained in more detail in the Methods section.
The results, presented in Table \ref{tab:att}, refer to the biotech sector between 1990 and 2010. We find that acquisitions significantly disrupt innovation in all four dimensions.

Specifically, the turnover of the inventors of the acquired company increases by 13.5\% in the five years following the acquisition compared to the turnover of the inventors not yet treated (i.e. the inventors who work at the target companies that will be acquired in the future).
After takeovers, we also find a significant increase in the number of inventors who no longer apply for patents (+6.3\%).
In terms of innovation productivity, takeovers lead to a significant decrease in the number of patents filed (-13.6\%).
If we look at the number of patents weighted by citations, the decline is even more pronounced (-35\%).

\begin{table}[ht]
    \centering
\begin{tabular}{lccc}
\toprule
 & ATT & Std. Error & 95\% Conf. Int. \\ 
\midrule
Left         & 0.135 & 0.016 & (0.10, 0.17) \\
R\&D Activity & -0.063& 0.009 & (-0.08, -0.05) \\
Log(patents)   & -0.136 & 0.027 & (-0.19, -0.08)\\
Log(citations) & -0.350 & 0.070 & (-0.49, -0.21)  \\
\midrule
\multicolumn{4}{l}{\footnotesize Control Group: Not Yet Treated; Estimation Method: Doubly Robust.} \\
\end{tabular}

        \caption{Average Treatment Effect on the Treated (ATT) after the acquisition}
        \label{tab:att}
\end{table}

\begin{figure}
    (a)\includegraphics[width=0.45\textwidth]{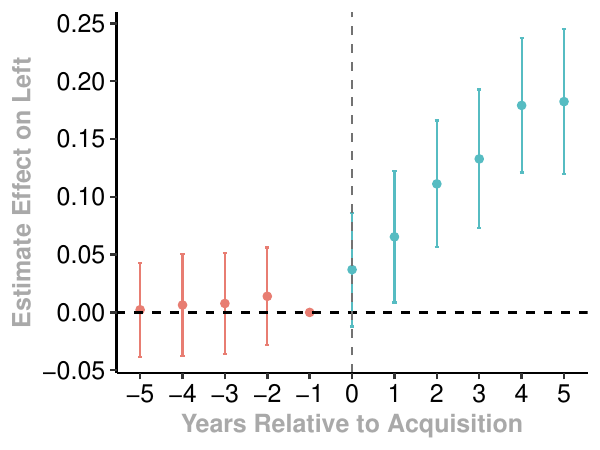} 
    (b)\includegraphics[width=0.45\textwidth]{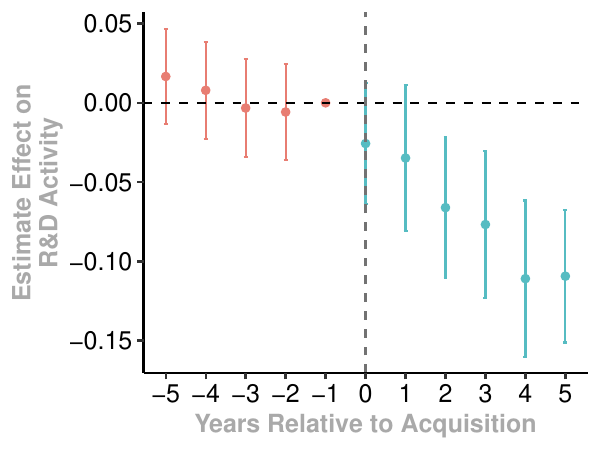} \\
     (c)\includegraphics[width=0.45\textwidth]{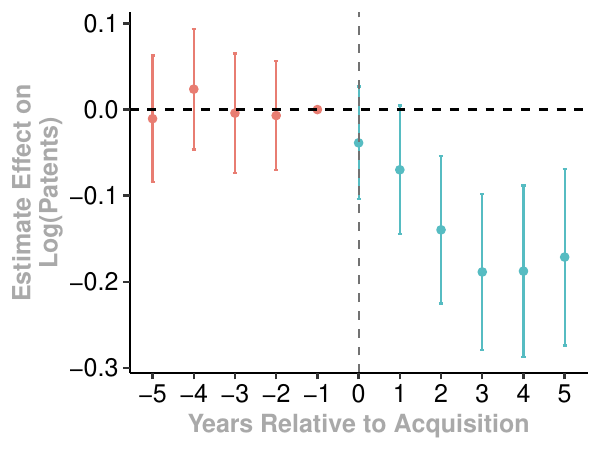}
     (d)\includegraphics[width=0.45\textwidth]{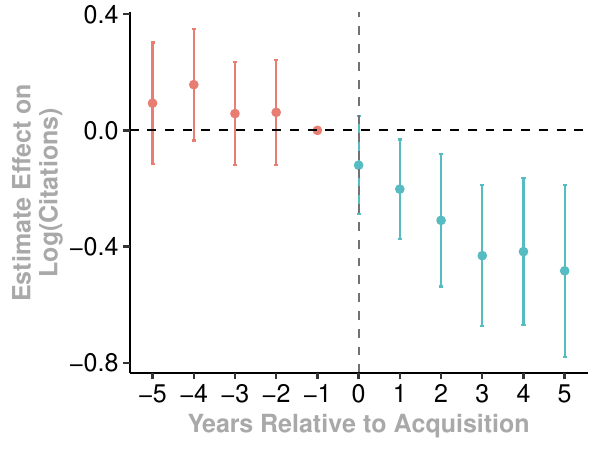}
    \caption{Average Treatment Effect on the Treated (ATT) before and after acquisitions. These figures show the estimated $\lambda_\tau$ from equation \ref{eq:event} together with the 95\% confidence interval. The year $t-1$ is the reference point. (a) examines the impact on inventor turnover. (b) looks at the probability that the inventor no longer applies for patents. (c) and (d) show the change in inventor productivity as measured by the number of patents and citations.}
    \label{fig:ATT}
\end{figure}

Figure \ref{fig:ATT} shows the effects of takeovers over the 5 years following the event. Since takeovers are not random events, the effects in the pre-acquisition phase could be anticipated to a certain extent.
In Figure \ref{fig:ATT}, however, we find no pre-trend for all dependent variables of interest.\footnote{As a further test, we perform the pre-test of the assumption of conditional parallel trends \citet{rios2022drdid}.In all cases, we find no evidence of pre-trends.}
In general, we find negative effects that become significant in the two years following the acquisition. The effects persist five years after the acquisition, with innovation productivity tending to level off after three years.

Using the staggered difference-in-differences analysis outlined in the Methods section, our study shows that acquisitions significantly affect acquired inventors.
We find inventors are significantly more likely to leave after an acquisition, suggesting that such corporate reorganisations create an environment of uncertainty.
This increase in departures can also be attributed to personnel changes such as layoffs or terminations that are part of post-acquisition phases.

The analysis goes beyond retention to highlight the impact on the survival of inventors in the industry following an acquisition. This aspect of our findings highlights the broader impact of acquisitions, not only on immediate employment status but also on long-term career trajectories.

In terms of productivity, our results suggest that both the number of patents filed and the impact of these patents, as measured by the number of citations, decline after the acquisition.
This reduction in innovation performance is concerning as it indicates a potential loss of innovation capacity following biotech takeovers.
Such a trend could be due to various factors, e.g. a change in corporate strategy, a change in workplace culture, a restructuring of R\&D or a reallocation of resources away from research and development activities.

These collective findings paint a picture of the acquisition event as a significant disruptive factor for inventors, which not only accelerates their departure --- whether voluntary or not --- but also has a detrimental effect on their innovation and productivity.
This observation raises important questions about the mechanisms that drive these patterns and how they can be mitigated to preserve human capital during corporate transitions.

In short, we find strong support for H1 that acquisitions negatively affect inventors' productivity with no clear signs of recovery in the five years following these events.

It is important to point out that our analysis is based on the assumption of common trends, which is validated by both visual inspection and statistical tests using the ``did'' package in R~\citep{did}. This validation strengthens the interpretations and conclusions we draw from our study.\footnote{For an alternative matched diff-in-diff approach, see \citep{verginer2022impact}. The results are consisted but a much smaller set of acquisition events is considered.}

\subsection*{Staying Versus Leaving}

To address H2, H3 and H4, which examine the impact of acquisitions on inventors' productivity as a function of whether they stayed in or left the firm, we estimate an aggregate after-effect as described in the Methods section, focusing on the total effect after the event.

Again, the overall effect in Table~\ref{tab:reg-base} shows a general decline in productivity after the takeover, as seen from the lower number of patents filed and the citations these patents received in the following years.
Notably, despite the general decline in productivity, the inventors who left the company before or after the takeover were more productive than those who stayed.

\begin{table}[h!]
    \centering
    \footnotesize
    \caption{Regression results, leave versus stay}\label{tab:reg-base}
    {
\def\sym#1{\ifmmode^{#1}\else\(^{#1}\)\fi}
\begin{tabular}{l*{6}{c}}
\toprule
                    &\multicolumn{1}{c}{(1)}&\multicolumn{1}{c}{(2)}&\multicolumn{1}{c}{(3)}&\multicolumn{1}{c}{(4)}&\multicolumn{1}{c}{(5)}&\multicolumn{1}{c}{(6)}\\
                    &\multicolumn{1}{c}{Left}&\multicolumn{1}{c}{log(Patents)}&\multicolumn{1}{c}{log(Citations)}&\multicolumn{1}{c}{Left}&\multicolumn{1}{c}{log(Patents)}&\multicolumn{1}{c}{log(Citations)}\\
\midrule
After               &       0.142\sym{*} &      -0.109\sym{*} &      -0.319\sym{*} &       0.115\sym{*} &      -0.126\sym{*} &      -0.351\sym{*} \\
                    &     (35.13)         &    (-12.28)         &    (-13.60)         &     (29.77)         &    (-14.53)         &    (-15.25)         \\
\addlinespace
Left                &                     &      -0.324\sym{*} &      -0.844\sym{*} &                     &      -0.344\sym{*} &      -0.879\sym{*} \\
                    &                     &    (-42.91)         &    (-41.31)         &                     &    (-44.61)         &    (-41.85)         \\
\addlinespace
Left $\times$ After &                     &       0.059\sym{*} &       0.219\sym{*} &                     &       0.056\sym{*} &       0.214\sym{*} \\
                    &                     &      (5.44)         &      (7.68)         &                     &      (5.28)         &      (7.59)         \\
\addlinespace
log(Tenure)         &                     &                     &                     &       0.178\sym{*} &       0.081\sym{*} &       0.251\sym{*} \\
                    &                     &                     &                     &     (70.45)         &     (13.22)         &     (15.64)         \\
\addlinespace
Exclusivity         &                     &                     &                     &      -0.222\sym{*} &      -0.084\sym{*} &      -0.149\sym{*} \\
                    &                     &                     &                     &    (-82.18)         &    (-14.96)         &     (-9.83)         \\
\addlinespace
log(Age)            &                     &                     &                     &       0.065\sym{*} &       0.010  &      -0.051\sym{*} \\
                    &                     &                     &                     &     (32.75)         &      (2.01)         &     (-3.79)         \\
\addlinespace
Common IPCs         &                     &                     &                     &      -0.008\sym{*} &       0.070\sym{*} &       0.161\sym{*} \\
                    &                     &                     &                     &    (-19.87)         &     (74.97)         &     (71.36)         \\
\addlinespace
Constant            &       0.228\sym{*} &       0.972\sym{*} &       2.638\sym{*} &      -0.120\sym{*} &       0.590\sym{*} &       1.776\sym{*} \\
                    &    (149.01)         &    (270.83)         &    (277.48)         &    (-29.06)         &     (63.25)         &     (72.25)         \\
\midrule
Acquiring Firm FE           &      Yes &   Yes  &       Yes &  Yes   &      Yes &  Yes    \\
Acquired Firm FE            &      Yes &   Yes  &       Yes &  Yes   &      Yes &  Yes    \\
Year FE                     &      Yes &   Yes  &       Yes &  Yes   &      Yes &  Yes    \\
\midrule
Observations        &      144\,532         &      144\,532         &      144\,532         &      144\,532         &      144\,532         &      144\,532         \\
\(R^{2}\)           &        0.32         &        0.11         &        0.21         &        0.40         &        0.15         &        0.24         \\
\textit{AIC}        &     120,592         &     345,942         &     630,043         &     102,645         &     338,108         &     624,297         \\
Log Lik.            &     -60,294         &    -172,967         &    -315,017         &     -51,317         &    -169,046         &    -312,140         \\
\bottomrule
\multicolumn{7}{l}{\tiny \textit{t} statistics in parentheses;  \sym{\dagger} \(p<0.01\), \sym{*} \(p<0.001\)}\\
\end{tabular}
}

\end{table}

After controlling for tenure, age, exclusivity, and common IPCs, we find that these factors also moderate productivity levels.
To better understand how these factors influence productivity as a function of the inventor's decision to stay or leave, we introduce an interaction term into the base model as described in the Methods Section.

In Table~\ref{tab:reg-age}, we investigate how age affects productivity after an acquisition, depending on whether the inventors stay in or leave the company.
Our results suggest that experienced inventors are more likely to stay with the company after the acquisition.
Experienced inventors who leave the company tend to experience a significant drop in productivity compared to their colleagues who stay with the company.
This suggests that experienced inventors are more productive if they remain in the company after the acquisition, which is consistent with H2. The acquiring company can better assess the value of experienced inventors whose record track of innovation productivity is more informative. The innovation potential of young acquired inventors is more uncertain. Therefore, they are more likely to leave the company after the acquisition and be more productive when they leave.  

\begin{table}[bh]
    \centering
    \small
        \caption{Regression results for Age}\label{tab:reg-age}
    {
\def\sym#1{\ifmmode^{#1}\else\(^{#1}\)\fi}
\begin{tabular}{l*{3}{cc}}
\toprule
                    &\multicolumn{2}{c}{(1)}           &\multicolumn{2}{c}{(2)}           &\multicolumn{2}{c}{(3)}           \\
                    &\multicolumn{2}{c}{Left}          &\multicolumn{2}{c}{log(Patents)}  &\multicolumn{2}{c}{log(Citations)}\\
\midrule
After               &       0.205\sym{*}&     (29.14)&      -0.243\sym{*}&    (-13.06)&      -0.736\sym{*}&    (-15.08)\\
log(Age)            &       0.083\sym{*}&     (40.78)&      -0.032\sym{*}&     (-5.45)&      -0.196\sym{*}&    (-12.56)\\
After $\times$ log(Age)&      -0.046\sym{*}&    (-13.91)&       0.060\sym{*}&      (6.43)&       0.198\sym{*}&      (8.52)\\
Left                &                     &            &      -0.896\sym{*}&    (-25.47)&      -2.464\sym{*}&    (-25.89)\\
Left $\times$ After &                     &            &       0.516\sym{*}&     (12.35)&       1.385\sym{*}&     (12.39)\\
Left $\times$ log(Age)&                     &            &       0.220\sym{*}&     (16.00)&       0.637\sym{*}&     (17.57)\\
Left $\times$ After $\times$ log(Age)&                     &            &      -0.177\sym{*}&    (-10.10)&      -0.442\sym{*}&     (-9.73)\\
\midrule
Exclusivity         &      -0.219\sym{*}&    (-81.68)&      -0.071\sym{*}&    (-12.38)&      -0.110\sym{*}&     (-7.17)\\
Common IPCs         &      -0.008\sym{*}&    (-19.63)&       0.070\sym{*}&     (75.19)&       0.162\sym{*}&     (71.70)\\
log(Tenure)         &       0.179\sym{*}&     (70.73)&       0.070\sym{*}&     (11.35)&       0.216\sym{*}&     (13.43)\\
Constant            &      -0.163\sym{*}&    (-38.84)&       0.694\sym{*}&     (61.01)&       2.131\sym{*}&     (69.07)\\
\midrule
Acquiring Firm FE            &      Yes &     &       Yes &     &      Yes &      \\
Acquired Firm FE            &      Yes &     &       Yes &     &      Yes &      \\
Year FE            &      Yes &     &       Yes &     &      Yes &      \\
\midrule
Observations        &      144\,532         &            &      144\,532         &            &      144\,532         &            \\
\(R^{2}\)           &        0.40         &            &        0.16         &            &        0.24         &            \\
\textit{AIC}        &     102,414         &            &     337,782         &            &     623,822         &            \\
Log Lik.            &     -51,200         &            &    -168,880         &            &    -311,900         &            \\
\bottomrule
\multicolumn{7}{l}{\tiny \textit{t} statistics in parentheses; \sym{\dagger} \(p<0.01\), \sym{*} \(p<0.001\)}\\
\end{tabular}
}

\end{table}

Shifting our focus to exclusivity --- a measure that reflects an inventor's external collaborations and their focus on firm-specific technologies -- we see some interesting patterns in Table~\ref{tab:reg-exclusivity}.
Inventors with higher exclusivity, i.e. those who primarily apply for patents only for the acquired company, are initially less likely to leave the company, and this does not change after the event, although the effect becomes less significant.

In terms of productivity, there is a significant drop in productivity for those with higher exclusivity, both in terms of the number of patents and the citation-weighted patent production after the event.
However, the productivity of those who leave the company after the event is higher than that of the similarly exclusive inventors who stay.
These observations suggest that acquiring firms retain inventors closely associated with the target company. However, on average, these inventors become more productive when they move to a third party. In line with H3, it is more challenging to evaluate innovations specific to the target company correctly. Difficulty in absorbing target firm-specific knowledge is one of the main causes of the decline in productivity in R\&D following takeovers.

\begin{table}
    \centering
    \small
        \caption{Regression results for Exclusivity}\label{tab:reg-exclusivity}
    {
\def\sym#1{\ifmmode^{#1}\else\(^{#1}\)\fi}
\begin{tabular}{l*{3}{cc}}
\toprule
                    &\multicolumn{2}{c}{(1)}           &\multicolumn{2}{c}{(2)}           &\multicolumn{2}{c}{(3)}           \\
                    &\multicolumn{2}{c}{Left}          &\multicolumn{2}{c}{log(Patents)}  &\multicolumn{2}{c}{log(Citations)}\\
\midrule
After               &       0.122\sym{*}&     (28.76)&      -0.107\sym{*}&    (-10.25)&      -0.282\sym{*}&    (-10.28)\\
Exclusivity         &      -0.217\sym{*}&    (-77.99)&      -0.030\sym{*}&     (-4.28)&       0.018         &      (0.94)\\
After $\times$ Exclusivity&      -0.017\sym{\dagger} &     (-3.26)&      -0.064\sym{*}&     (-4.59)&      -0.218\sym{*}&     (-6.16)\\
Left                &                     &            &      -0.245\sym{*}&    (-24.72)&      -0.584\sym{*}&    (-22.06)\\
Left $\times$ After &                     &            &      -0.039\sym{\dagger} &     (-2.78)&      -0.082  &     (-2.22)\\
Left $\times$ Exclusivity&                     &            &      -0.317\sym{*}&    (-18.41)&      -0.940\sym{*}&    (-19.52)\\
Left $\times$ After $\times$ Exclusivity&                     &            &       0.300\sym{*}&     (12.40)&       0.918\sym{*}&     (14.16)\\
\midrule
Common IPCs         &      -0.008\sym{*}&    (-19.85)&       0.070\sym{*}&     (74.99)&       0.161\sym{*}&     (71.39)\\
log(Tenure)         &       0.178\sym{*}&     (70.82)&       0.067\sym{*}&     (10.84)&       0.211\sym{*}&     (13.02)\\
log(Age)            &       0.066\sym{*}&     (32.92)&       0.014\sym{\dagger} &      (2.78)&      -0.039\sym{\dagger} &     (-2.84)\\
Constant            &      -0.124\sym{*}&    (-30.53)&       0.583\sym{*}&     (60.33)&       1.750\sym{*}&     (68.29)\\
\midrule
Acquiring Firm FE            &      Yes &     &       Yes &     &      Yes &      \\
Acquired Firm FE            &      Yes &     &       Yes &     &      Yes &      \\
Year FE            &      Yes &     &       Yes &     &      Yes &      \\
\midrule
Observations        &      144\,532         &            &      144\,532         &            &      144\,532         &            \\
\(R^{2}\)           &        0.40         &            &        0.16         &            &        0.24         &            \\
\textit{AIC}        &     102,633         &            &     337,834         &            &     623,957         &            \\
Log Lik.            &     -51,310         &            &    -168,906         &            &    -311,968         &            \\
\bottomrule
\multicolumn{7}{l}{\tiny \textit{t} statistics in parentheses; \sym{\dagger} \(p<0.01\), \sym{*} \(p<0.001\)}\\
\end{tabular}
}

\end{table}

This observation prompts us to examine the role of technological overlap in shaping productivity outcomes more closely. Concerning  H4, we examine whether the extent of technological fit between an inventor and the acquiring firm affects productivity and the likelihood of retention. Our analysis focuses on the impact of common International Patent Classification (IPC) subgroup classes between the inventor and the acquiring firm on these metrics.

\begin{table}[ht]
\centering
\small
    \caption{Regression results for IPC Overlap}\label{tab:reg-ipc}
    {
\def\sym#1{\ifmmode^{#1}\else\(^{#1}\)\fi}
\begin{tabular}{l*{3}{cc}}
\toprule
                    &\multicolumn{2}{c}{(1)}           &\multicolumn{2}{c}{(2)}           &\multicolumn{2}{c}{(3)}           \\
                    &\multicolumn{2}{c}{Left}          &\multicolumn{2}{c}{log(Patents)}  &\multicolumn{2}{c}{log(Citations)}\\
\midrule
Before $\times$ 0 IPC&       0.000         &          &                     &            &                     &            \\
Before $\times$ 1-3 IPC&      -0.111\sym{*}&    (-22.38)&                     &            &                     &            \\
Before $\times$ 4+ IPC&      -0.150\sym{*}&    (-28.34)&                     &            &                     &            \\
After $\times$ 0 IPC&       0.126\sym{*}&     (15.57)&                     &            &                     &            \\
After $\times$ 1-3 IPC&      -0.003         &     (-0.39)&                     &            &                     &            \\
After $\times$ 4+ IPC&      -0.030\sym{*}&     (-4.45)&                     &            &                     &            \\
Stayed $\times$ Before $\times$ 0 IPC&                     &            &       0.000         &         (.)&       0.000         &         (.)\\
Stayed $\times$ Before $\times$ 1-3 IPC&                     &            &       0.096\sym{*}&      (8.78)&       0.182\sym{*}&      (5.73)\\
Stayed $\times$ Before $\times$ 4+ IPC&                     &            &       0.459\sym{*}&     (38.52)&       1.042\sym{*}&     (30.84)\\
Stayed $\times$ After $\times$ 0 IPC&                     &            &      -0.092\sym{*}&     (-4.08)&      -0.294\sym{*}&     (-5.02)\\
Stayed $\times$ After $\times$ 1-3 IPC&                     &            &       0.025         &      (1.71)&      -0.038         &     (-0.94)\\
Stayed $\times$ After $\times$ 4+ IPC&                     &            &       0.280\sym{*}&     (18.20)&       0.578\sym{*}&     (13.82)\\
Left $\times$ Before $\times$ 0 IPC&                     &            &      -0.359\sym{*}&    (-23.13)&      -0.989\sym{*}&    (-22.52)\\
Left $\times$ Before $\times$ 1-3 IPC&                     &            &      -0.291\sym{*}&    (-21.23)&      -0.815\sym{*}&    (-20.81)\\
Left $\times$ Before $\times$ 4+ IPC&                     &            &       0.176\sym{*}&     (10.15)&       0.384\sym{*}&      (8.23)\\
Left $\times$ After $\times$ 0 IPC&                     &            &      -0.310\sym{*}&    (-16.47)&      -0.707\sym{*}&    (-13.17)\\
Left $\times$ After $\times$ 1-3 IPC&                     &            &      -0.288\sym{*}&    (-19.27)&      -0.836\sym{*}&    (-19.80)\\
Left $\times$ After $\times$ 4+ IPC&                     &            &      -0.012         &     (-0.73)&      -0.044         &     (-1.01)\\
\midrule
Exclusivity         &      -0.218\sym{*}&    (-80.57)&      -0.080\sym{*}&    (-13.97)&      -0.138\sym{*}&     (-9.05)\\
log(Tenure)         &       0.174\sym{*}&     (68.74)&       0.080\sym{*}&     (12.90)&       0.252\sym{*}&     (15.49)\\
log(Age)            &       0.065\sym{*}&     (32.49)&       0.030\sym{*}&      (5.73)&      -0.011         &     (-0.79)\\
Constant            &      -0.026\sym{*}&     (-4.23)&       0.567\sym{*}&     (42.42)&       1.755\sym{*}&     (46.84)\\
\midrule
Acquiring Firm FE            &      Yes &     &       Yes &     &      Yes &      \\
Acquired Firm FE            &      Yes &     &       Yes &     &      Yes &      \\
Year FE            &      Yes &     &       Yes &     &      Yes &      \\
\midrule
Observations        &      144\,532         &            &      144\,532         &            &      144\,532         &            \\
\(R^{2}\)           &        0.40         &            &        0.14         &            &        0.23         &            \\
\textit{AIC}        &     102,003         &            &     340,888         &            &     625,939         &            \\
Log Lik.            &     -50,992         &            &    -170,429         &            &    -312,954         &            \\
\bottomrule
\multicolumn{7}{l}{\tiny \textit{t} statistics in parentheses; \sym{\dagger} \(p<0.01\), \sym{*} \(p<0.001\)}\\
\end{tabular}
}

\end{table}

Table~\ref{tab:reg-ipc} shows how technological overlap, as measured by common International Patent Classification (IPC) subgroup classes, affects inventors' retention and productivity after acquisition.
The results indicate that inventors with a significant overlap (4 or more common IPC classes) with the acquiring company are less likely to exit after the acquisition. In contrast, inventors without a common IPC are significantly more likely to leave.
This trend supports the idea that acquiring companies retain inventors whose expertise is better aligned with their ongoing R\&D activities.\footnote{Similar results are obtained when focusing on the most important acquisitions and use a matched diff-in-diff regression strategy \citep{verginer2022impact}. A breakdown with five levels of similarity, showing qualitatively identical results, is available in the appendix in Table~\ref{tab:extended-ipc}}.

The analysis also shows that post-acquisition, inventors with the highest degree of technological overlap (4 or more IPC classes) experience the most significant increase in productivity, while those with no or partial (0 and 1-3 IPCs in common) experience a drop in productivity, regardless of whether they stayed or not. 

To summarize, inventors with strong technological ties to the acquiring company are less likely to leave the company and inventors without technological overlaps suffer a more significant decline in productivity if they stay with the acquiring company.
Inventors whose expertise closely matches the current R\&D focus of the acquiring company not only tend to stay, but also experience increased productivity.
Conversely, inventors with little or no overlap in technological areas find it challenging to find a good fit inside and outside the acquiring company.
Taken together, these results support hypothesis 4, that technological overlaps facilitate the absorption of the target company's knowledge. Therefore, the disruption of innovation productivity is much stronger for acquired inventors without technological overlaps.

The combined effects of age, exclusivity, and technological similarity on inventor productivity post-acquisition reveal the complex interplay of talent retention and innovation.
Older inventors, with their experience and established expertise, tend to stay with the acquiring firm and maintain their productivity.
In contrast, exclusive inventors, despite their specialised knowledge of the acquired firm's technology, may struggle to assimilate into the new environment and, therefore, become more productive elsewhere.
In addition, inventors with a technological overlap above the median are more likely to remain in the acquiring firm and experience higher innovation productivity. This suggests that a low or no technological match could lead to underutilising their skills in the new organizational environment.
Overall, the decline in inventor productivity following biotech acquisitions is driven by fundamental uncertainty and disagreement in assessing the potential value of young and exclusive inventors. Even among inventors working in different technology areas, there is a significant decline in productivity following acquisitions.

\section{Conclusion}\label{sec:conclusion}
Firm acquisition is one possible solution for sourcing external knowledge when technological change is fast-paced and knowledge is complex, such as in the biotechnology industry \citep{carayannopoulos2010}.
However, to absorb new knowledge, the activities of the newly combined firms must be reorganised \citep{capron2001,karim2000,colombo2014}.
The reorganisation of R\&D activities can translate into uncertainty and conflicts in the aftermath of the acquisition, and the creative labour force might decide to leave, hampering the potential benefits of the acquisition \citep{Ernst2000,paruchuri2006,Kapoor2007,Hussinger2018,arroyabe2020,Kim2022}.

As noticed in \citet{colombo2014} among others, the analysis of the impact of the acquisition on R\&D performance and inventors' departure is limited by causal ambiguity.
Furthermore, few studies take the inventors' perspective, as it is challenging to track their innovation productivity at the individual level over a sufficiently long period after the acquisition \citep{Kapoor2007}.
Therefore, in this paper, we apply a generalised difference-in-differences approach to fill this gap. 

In particular, we focus on the takeover wave of biotech firms at the turn of the century, a well-known example of takeovers to absorb new knowledge. We use a new dataset that disambiguates inventors and assignees to track patent production over a long period \citep{Morrison2017}. We take the perspective of biotech innovators to analyse the impact of acquisitions on their R\&D productivity. We document a significant and persistent disruption in innovation productivity: inventors are much more likely to leave (+13.5\%) or become inactive (+6.3\%) after takeovers. In general, patent production is considerably lower (-13.6\% patents, -35\% citation-weighted patents).
Our findings are robust to various robustness checks, and the magnitude of turnover we find is in line with \citet{Ernst2000}, \citet{Ranft2000} and \citet{Carriquiry2017}.

Our results paint a complex picture of turnover dynamics after an acquisition. On the one hand, turnover is, as expected, higher immediately after an acquisition, which is consistent with previous findings. On the other hand, some inventors that we expect to be the most valuable are also the most likely to leave, leading to a significant decrease in productivity among inventors who remain with the acquiring firm.

Considering that R\&D employees are considered particularly important by acquiring firms themselves \citep{Ranft2000}, our results suggest that a causal channel for the reduction in R\&D performance after acquisitions is the inherent uncertainty and disagreement about the potential future impact of the acquired inventors' research ideas. If the acquired inventors are young, it is difficult to determine their future innovation potential. Therefore, young inventors who stay have more difficulty adapting to the new organizational environment and are more productive when they leave. The situation is similar for inventors whose production is more specific to the target company: they have more difficulty integrating into the acquiring company and are more productive outside it.
Therefore, an acquisition can potentially diminish the value of the acquired target and hamper post-acquisition integration and success.

Our findings, taken together, suggest that if an acquisition target has inventors working exclusively there, their retention is more likely.
Still, an increased turnover rate should be expected, especially among inventors with more dissimilar knowledge.

With this study, we contribute to the literature on post-merger integration by providing additional evidence that acquisitions accelerate the departure of inventors, which entails huge social costs in terms of lower innovator productivity, especially for inventors who do not leave the acquiring firm. Apart from the situations in which the acquiring firm voluntarily destroys innovations \citep{cunningham2021}, we suspect that most of the disruption is involuntary.
The effect is mitigated by the technological similarity between the inventors and the acquiring firm and by the experience of the inventors. Acquired inventors who have filed patents with multiple assignees can also better cope with the organizational change to collaborate with the acquiring company.

However, asymmetric information and uncertainty about the potential value of R\%D projects imply inefficient allocation of resources. In addition, the disruption of research teams in the target company and the reconfiguration of these teams for integration into the acquiring company are likely to play a role in reducing productivity. Future research can be devoted to investigating the effects of team disruption and reconfiguration on R\&D productivity after acquisitions to find viable strategies to mitigate these effects.

On another level, the increasing availability of matched employer/employee data could help shed more light on the decision to leave the acquiring organization. Some of these datasets \cite{arnold2023job} collect information on voluntary terminations and involuntary departures, which could clarify who is responsible for inventor turnover. Finally, the relationship between innovation and competition deserves more attention to prevent killer acquisitions that cause even higher costs by restricting competition in which the acquiring firm voluntarily destroys innovations \citep{cunningham2021}; we suspect that most of the disruption is involuntary.

In conclusion, our study underscores the critical, yet often underappreciated, consequences of firm acquisitions on the turnover and productivity of R\&D personnel, particularly in the biotech sector. While acquisitions are strategic moves to assimilate novel knowledge, our findings reveal that they frequently result in significant disruptions, leading to increased turnover and reduced productivity among inventors. This turnover not only reflects the challenges of post-acquisition integration but also highlights the potential for misalignment of expectations and objectives between the acquiring and acquired firms.

\setstretch{1} %
\bibliographystyle{apalike} %
\bibliography{turnover.bib} %

\clearpage
\appendix
\section*{Appendix}

\begin{table}[ht]
    \caption{Descritpives statistics for main variables}\label{tab:descriptives}
    \centering
\begin{tabular}{lrrrrrr}
\toprule
Variable & Mean & SD & Median & Min & Max & N\\
\midrule
Patents & 2.69 & 11.91 & 1 & 1 & 3\,926 & 15\,318\\
Citations & 223.78 & 1\,462.45 & 7 & 0 & 77\,875 & 15,318\\
Age & 8.40 & 7.34 & 6 & 2 & 65 & 15\,318\\
Tenure & 5.32 & 2.31 & 4 & 2 & 23 & 15\,318\\
Common IPC & 3.24 & 2.93 & 3 & 0 & 38 & 15\,318\\
Exclusivity & 0.44 & 0.45 & 0.29 & 0 & 1 & 15\,318\\
Deal Year & 2002.3 & 5.05 & 2003 & 1990 & 2010 & 15\,318\\
\bottomrule
\end{tabular}

\end{table}

\begin{table}[ht]
    \small
    \caption{Correlations of main variables}\label{tab:corr}
    \begin{tabular}{lrrrrrr}
\toprule
{}            &Common IPC& Citations&  Patents &  Exclusivity &  Tenure &  Age \\
\midrule
Common IPC     &   1 &          &          &              &         &       \\
Citations      &   0.064 &    1 &          &              &         &       \\
Patents        &   0.076 &    0.121 &    1 &              &         &       \\
Exclusivity    &   0.024 &   -0.040 &   -0.018 &        1 &         &       \\
Tenure         &   0.127 &   -0.015 &    0.003 &        0.248 &   1 &       \\
Age            &   0.018 &    0.025 &    0.031 &        0.002 &   0.296 &   1 \\
\bottomrule
\end{tabular}

\end{table}
\begin{table}[]
    \small
    \caption{Extended IPC Regression with 5 levels}\label{tab:extended-ipc}
    {
\def\sym#1{\ifmmode^{#1}\else\(^{#1}\)\fi}
\begin{tabular}{l*{3}{cc}}
\toprule
                    &\multicolumn{2}{c}{(1)}           &\multicolumn{2}{c}{(2)}           &\multicolumn{2}{c}{(3)}           \\
                    &\multicolumn{2}{c}{Left}          &\multicolumn{2}{c}{log(Patents)}  &\multicolumn{2}{c}{log(Citations)}\\
\midrule
Before $\times$ 0 IPC&       0.000         &         (.)&                     &            &                     &            \\
Before $\times$ 1 IPC&      -0.086\sym{*}&    (-16.31)&                     &            &                     &            \\
Before $\times$ 2 IPC&      -0.118\sym{*}&    (-21.85)&                     &            &                     &            \\
Before $\times$ 3 IPC&      -0.152\sym{*}&    (-27.26)&                     &            &                     &            \\
Before $\times$ 4+ IPC&      -0.161\sym{*}&    (-30.24)&                     &            &                     &            \\
After $\times$ 0 IPC&       0.125\sym{*}&     (15.45)&                     &            &                     &            \\
After $\times$ 1 IPC&       0.021\sym{\dagger} &      (2.76)&                     &            &                     &            \\
After $\times$ 2 IPC&      -0.005         &     (-0.70)&                     &            &                     &            \\
After $\times$ 3 IPC&      -0.046\sym{*}&     (-5.71)&                     &            &                     &            \\
After $\times$ 4+ IPC&      -0.042\sym{*}&     (-6.22)&                     &            &                     &            \\
Stayed $\times$ Before $\times$ 0 IPC&                     &            &       0.000         &         (.)&       0.000         &         (.)\\
Stayed $\times$ Before $\times$ 1 IPC&                     &            &       0.038\sym{\dagger} &      (3.14)&       0.027         &      (0.77)\\
Stayed $\times$ Before $\times$ 2 IPC&                     &            &       0.097\sym{*}&      (7.87)&       0.192\sym{*}&      (5.38)\\
Stayed $\times$ Before $\times$ 3 IPC&                     &            &       0.215\sym{*}&     (16.57)&       0.484\sym{*}&     (13.06)\\
Stayed $\times$ Before $\times$ 4+ IPC&                     &            &       0.489\sym{*}&     (40.57)&       1.119\sym{*}&     (32.83)\\
Stayed $\times$ After $\times$ 0 IPC&                     &            &      -0.089\sym{*}&     (-3.93)&      -0.286\sym{*}&     (-4.87)\\
Stayed $\times$ After $\times$ 1 IPC&                     &            &      -0.054\sym{\dagger} &     (-2.90)&      -0.202\sym{*}&     (-3.97)\\
Stayed $\times$ After $\times$ 2 IPC&                     &            &       0.059\sym{\dagger} &      (3.25)&       0.029         &      (0.57)\\
Stayed $\times$ After $\times$ 3 IPC&                     &            &       0.120\sym{*}&      (6.24)&       0.191\sym{*}&      (3.76)\\
Stayed $\times$ After $\times$ 4+ IPC&                     &            &       0.312\sym{*}&     (20.12)&       0.659\sym{*}&     (15.66)\\
Left $\times$ Before $\times$ 0 IPC&                     &            &      -0.353\sym{*}&    (-22.70)&      -0.974\sym{*}&    (-22.13)\\
Left $\times$ Before $\times$ 1 IPC&                     &            &      -0.362\sym{*}&    (-21.65)&      -1.003\sym{*}&    (-20.83)\\
Left $\times$ Before $\times$ 2 IPC&                     &            &      -0.268\sym{*}&    (-14.86)&      -0.773\sym{*}&    (-14.97)\\
Left $\times$ Before $\times$ 3 IPC&                     &            &      -0.142\sym{*}&     (-6.93)&      -0.404\sym{*}&     (-7.10)\\
Left $\times$ Before $\times$ 4+ IPC&                     &            &       0.207\sym{*}&     (11.86)&       0.463\sym{*}&      (9.89)\\
Left $\times$ After $\times$ 0 IPC&                     &            &      -0.302\sym{*}&    (-16.00)&      -0.685\sym{*}&    (-12.75)\\
Left $\times$ After $\times$ 1 IPC&                     &            &      -0.300\sym{*}&    (-16.35)&      -0.853\sym{*}&    (-16.34)\\
Left $\times$ After $\times$ 2 IPC&                     &            &      -0.285\sym{*}&    (-15.88)&      -0.879\sym{*}&    (-17.28)\\
Left $\times$ After $\times$ 3 IPC&                     &            &      -0.207\sym{*}&    (-10.59)&      -0.584\sym{*}&    (-10.97)\\
Left $\times$ After $\times$ 4+ IPC&                     &            &       0.023         &      (1.41)&       0.044         &      (1.01)\\
Exclusivity         &      -0.217\sym{*}&    (-80.26)&      -0.081\sym{*}&    (-14.25)&      -0.142\sym{*}&     (-9.32)\\
log(Tenure)         &       0.172\sym{*}&     (67.95)&       0.084\sym{*}&     (13.45)&       0.261\sym{*}&     (16.06)\\
log(Age)            &       0.066\sym{*}&     (33.16)&       0.026\sym{*}&      (5.01)&      -0.021         &     (-1.52)\\
Constant            &      -0.017\sym{\dagger} &     (-2.86)&       0.548\sym{*}&     (40.75)&       1.706\sym{*}&     (45.27)\\
\midrule
Acquiring Firm FE            &      Yes &     &       Yes &     &      Yes &      \\
Acquired Firm FE            &      Yes &     &       Yes &     &      Yes &      \\
Year FE            &      Yes &     &       Yes &     &      Yes &      \\
\midrule
Observations        &      144\,532         &            &      144\,532         &            &      144\,532         &            \\
\(R^{2}\)           &        0.40         &            &        0.14         &            &        0.23         &            \\
\textit{AIC}        &     101,696         &            &     340,483         &            &     625,561         &            \\
Log Lik.            &     -50,835         &            &    -170,218         &            &    -312,757         &            \\
\bottomrule
\multicolumn{7}{l}{\footnotesize \textit{t} statistics in parentheses}\\
\multicolumn{7}{l}{\footnotesize  \(p<0.05\), \sym{\dagger} \(p<0.01\), \sym{*} \(p<0.001\)}\\
\end{tabular}
}

\end{table}

\end{document}